\documentclass[12pt]{iopart}

\expandafter\let\csname equation*\endcsname\relax
\expandafter\let\csname endequation*\endcsname\relax 

\usepackage[english]{babel}
\usepackage{amsmath}
\usepackage{amssymb}
\usepackage{graphicx}
\usepackage{color}

\usepackage{graphicx}
\usepackage{caption}
\usepackage{subcaption}

\usepackage[utf8]{inputenc}
\usepackage{amsmath}
\usepackage{amsfonts}
\usepackage{graphicx} 
\usepackage{amsthm}
\usepackage{placeins}
\usepackage{verbatim}
\usepackage{booktabs}
\usepackage{cite}
\usepackage{color}
\usepackage{bm}
\usepackage{bbold}
\usepackage{nicefrac}
\usepackage[colorlinks=true,linkcolor=red,citecolor=blue]{hyperref}
\usepackage{comment}
\usepackage[all]{xy}
\usepackage[width=17.5cm, vscale=0.8]{geometry}
\usepackage{booktabs}

\newcommand{\beq}{\begin{equation}}
\newcommand{\eeq}{\end{equation}}
\newcommand{\be}{\begin{equation}}
\newcommand{\ee}{\end{equation}}

\newcommand{\dd}{\mathrm{d}}

\renewcommand{\e}{\mathrm{e}}

\begin{document}

\title[Large deviations of the maximum of i.i.d. random variables]{Large deviations of the maximum of independent and identically distributed random variables}

\author{Pierpaolo Vivo}
\address{King's College London, Department of Mathematics, Strand, London WC2R 2LS, United Kingdom}

\date{\today}

\begin{abstract}
A pedagogical account of some aspects of Extreme Value Statistics (EVS) is presented from the somewhat non-standard viewpoint of Large Deviation Theory. We address the following problem: given a set of $N$ i.i.d. random variables $\{X_1,\ldots,X_N\}$ drawn from a parent probability density function (pdf) $p(x)$, what is the probability that the maximum value of the set $X_{\mathrm{max}}=\max_i X_i$ is ``atypically larger" than expected? The cases of exponential and Gaussian distributed variables are worked out in detail, and the right rate function for a general pdf in the Gumbel basin of attraction is derived. The Gaussian case convincingly demonstrates that the full rate function \emph{cannot} be determined from the knowledge of the limiting distribution (Gumbel) \emph{alone}, thus implying that it indeed carries additional information. Given the simplicity and richness of the result and its derivation, its absence from textbooks, tutorials and lecture notes on EVS for physicists appears inexplicable.

\end{abstract}

\maketitle


\section{Introduction}

Extreme Value Statistics (EVS) and Large Deviations Theory (LDT) are undoubtedly among the most solid and fertile theoretical masterpieces of modern probability theory. Developed independently over the course of several decades by top-class mathematicians, they have both gradually percolated into the domain of Statistical Physics (SP), to the extent that LDT is now recognized as \emph{the} proper language in which SP formalism should be expressed, and cutting-edge research on the EVS of correlated variables is nowadays the bread and butter of dozens of colleagues.

At odds with the widespread impact LDT and EVS have produced outside the realm of rigorous mathematics, physicists have been somehow reluctant to put together truly accessible and pedagogical accounts of their fundamentals, with the exception of highly commendable but isolated enterprises (see e.g. \cite{ellis,touchette1,touchette2} for LDT - \cite{arnold,majumdarpal,majumdarschehr,fortin,biroli} for EVS - and references therein). One of the unfortunate consequences is that neither theory is typically taught or integrated in standard physics curricula around the globe.

In their ``classical" (textbook) descriptions, EVS primarily deals with (among other observables) the statistics of the \emph{maximum} $X_{\mathrm{max}}$ (or \emph{minimum}) of a set of random variables $\{X_1,\ldots,X_N\}$, while LDT is concerned with atypical fluctuations of a random variable $S_N$ (depending on a parameter $N$) away from its expected value $\langle S_N\rangle$, which decay \emph{exponentially} fast as the parameter $N$ increases. LDT estimates are typically written in the form
\be
\mathrm{Prob}[S_N\leq s]\approx
\begin{cases}
\exp\left(-\omega_N^{(\ell)}\psi_\ell(s)\right)\ ,&\qquad s<\langle S_N\rangle\label{LDTdef}\\
1-\exp\left(-\omega_N^{(r)}\psi_r(s)\right)\ ,&\qquad s>\langle S_N\rangle\ ,
\end{cases}
\ee
where the nonzero \emph{left} and \emph{right rate functions} $\psi_{\ell,r}(s)$ control the (exponentially small) probability that $S_N$ takes values anomalously smaller or larger than $\langle S_N\rangle$, respectively. The symbol $\approx$ in \eqref{LDTdef} stands for $\lim_{N\to\infty}-\ln \mathrm{Prob}[S_N\leq s]/\omega_N^{(\ell)}=\psi_\ell(s)$ and similarly on the right. Note that nontrivial limits $\psi_{\ell,r}(s)$ can only be obtained by tuning the \emph{speeds} $\omega_N^{(\ell)}$ and $\omega_N^{(r)}$ of the large deviation estimate to precise functions of $N$. As an example of this formalism, $S_N$ may be taken to be the \emph{sample mean} $S_N=(1/N)\sum_{i=1}^N Y_i$ of independent and identically distributed (i.i.d.) random variables $\{Y_1,\ldots,Y_N\}$, drawn from a common parent probability density function (pdf), see \cite{touchette1} for a set of instructive examples worked out in detail.

From the exceedingly concise summary in the last paragraph, it is hard to speculate whether a connection between EVS and LDT should exist at all. They simply seem to target different attributes: ``big" \emph{vs.} ``anomalously rare". However, a moment of reflection should induce a quite natural question: what if the random variable $S_N$ (subject to atypical fluctuations) is taken to be $X_{\mathrm{max}}$ itself\footnote{Obviously, $X_{\mathrm{max}}$ depends on the sample size $N$.}, instead of the sample mean of the $X_i$'s? In other words, what is the probability that the \emph{maximum} of a set of random variables is ``atypically larger" (or smaller) than its expected value? 

Problems of this ilk have been addressed at length in the context of a certain type of \emph{strongly correlated} random variables, namely the eigenvalues of random matrices (see \cite{majumdarreview} and references therein). It felt just natural to assume that the problem for i.i.d. random variables (\emph{a priori} simpler) must have been settled long before.

Much to my surprise, I was able to retrieve only a single paper \cite{giuliano} where the EVS of i.i.d. random variables was looked at through the prism of LDT. The authors of \cite{giuliano} must have felt the same bewilderment as they wrote ``We are not aware of any other work on extreme value theory with
results formulated in this way.". However, the formal style and the intended audience of \cite{giuliano} make it a tough reading for the uninitiated.
 
I will argue here that this problem is at the same time rich, instructive and particularly simple (yet nontrivial) to deserve to be analyzed in detail and presented in a form accessible to an audience of trained physicists. This will be done by first introducing some preliminary notions (often not easy to find elsewhere) on ``classical" EVS, keeping the style as informal as possible.

\section{Preliminaries on ``classical" EVS for i.i.d. random variables}

Consider a collection of $N$ i.i.d. random variables $\{X_1,\ldots,X_N\}$, all drawn from the same continuous pdf
$p(x)$. We denote by $P(x)$ the cumulative distribution function (cdf) of each of the $X_i$'s, $P(x)=\int^x\dd y\ p(y)$. Also, we denote the maximum of the set $\{X_i\}$ by $X_{\mathrm{max}}=\max_i \{X_i\}$.

The cdf of $X_{\mathrm{max}}$ (denoted in the following by $Q_N(x)$) can be easily written as
\be
Q_N(x)=\mathrm{Prob}[X_{\mathrm{max}}\leq x]=\int^x\cdots\int^x \dd x_1\cdots \dd x_N p(x_1)\cdots p(x_N)=\left[\underbrace{\int^x\dd y p(y)}_{P(x)}\right]^N\ ,
\ee
where one uses the fact that the maximum is smaller than $x$ only if each of the variables is, and the independence of the variables. 

What happens now for $N\to\infty$? It is clear that $\lim_{N\to\infty}Q_N(x)$ for $x$ fixed is disappointingly trivial: since $0\leq P(x)\leq 1$,
the limit of $P(x)^N$ can only take two possible values: $0$ or $1$. In order to obtain a nontrivial limiting distribution, one has to send \emph{both} $N,x\to\infty$, in such a way that the combination $z=(x-a_N)/b_N$ is kept constant for suitably chosen \emph{centering} and \emph{scaling} constants $a_N\in\mathbb{R}$ and $b_N>0$, respectively. 

The standard goal of classical EVS can be summarized as follows: find $a_N$, $b_N$ and $F(z)$ (the latter independent of $N$) such that
\be
\lim_{N\to\infty}Q_N(a_N+b_N z)=F(z)\ .
\ee

The celebrated Fisher-Tippett-Gnedenko theorem \cite{fishertippett,gumbel,gnedenko} states that $F(z)$ can \emph{only} be of \emph{three} different types (Gumbel, Fr\'echet and Weibull), depending on the right tail of the parent pdf $p(x)$. Informally, if  we denote by $x^\star=\sup(x:P(x)<1)$ the upper endpoint of the support of $p(x)$

\begin{figure}[htb]
\begin{center}
\includegraphics[totalheight=0.17\textheight]{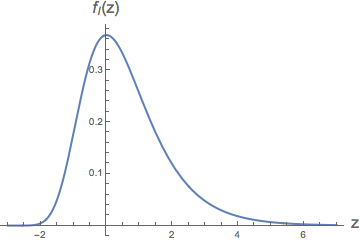}
\includegraphics[totalheight=0.17\textheight]{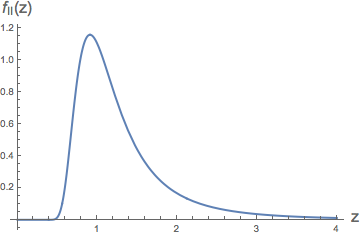}
\includegraphics[totalheight=0.17\textheight]{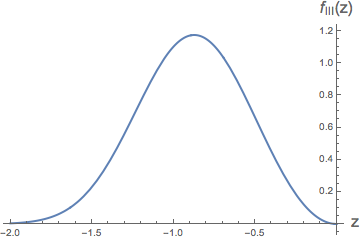}
\caption{Left to right: the pdfs $f_{\mathrm{I}}(z)$ (Gumbel), $f_{\mathrm{II}}(z)$ (Fr\'echet) and $f_{\mathrm{III}}(z)$ (Weibull).\label{densityfigs}}
\end{center}
\end{figure}

\begin{itemize}
\item If $x^\star$ is finite or infinite, and $p(x)$ falls off faster than any power for $x\to x^\star$ (for instance in the exponential and Gaussian cases), then the limiting distribution $F(z)$ is {\em Gumbel}, $F_{\mathrm{I}}(z)=\exp(-\exp(-z))$. 
\item If $x^\star$ is infinite and $p(x)$ falls off as a power law, $p(x)\sim x^{-(\gamma+1)}$, then the limiting distribution $F(z)$ is {\em Fr\'echet}, $F_{\mathrm{II}}(z)=\e^{-1/z^\gamma}$ if $z> 0$ and $0$ otherwise.
\item If $x^\star$ is finite, for instance $p(x)=0$ for $x>1$ and $p(x)\sim (1-x)^{\gamma-1}$ when $x\to 1^-$ with $\gamma>0$, then
the limiting distribution $F(z)$ is {\em Weibull}, $F_{\mathrm{III}}(z)=\e^{-|z|^\gamma}$ for $z<0$ and $1$ otherwise.
\end{itemize}
A more formal classification of basins of attraction can be found in \cite{ferreira}, Theorem 1.2.1. In Fig. \ref{densityfigs} I plot the pdfs corresponding to the three classes above.

The tail cumulative distribution function $\tilde{Q}_N(x)=\mathrm{Prob}[X_{\mathrm{max}}> x]=1-Q_N(x)$ satisfies obviously
\be
\lim_{N\to\infty}\tilde{Q}_N(a_N+b_N z)=1-F(z)\ ,
\ee
with the same constants $a_N$ and $b_N$.

I summarize here three results \cite{salvadori} that are all of practical importance, but hard to find simultaneously stated on the same page.
\begin{enumerate}
\item The constants $a_N$ and $b_N$ can be found as follows ($P^{-1}(x)$ denotes the functional inverse of the cdf $P(x)$, if expressible in a closed form)\\
(a) Gumbel\\
\be
a_N=P^{-1}\left(1-\frac{1}{N}\right)\mbox{  and  }b_N=P^{-1}\left(1-\frac{1}{N\mathrm{e}}\right)-a_N\ .\label{ANgumbel}
\ee
(b) Fr\'echet
\be
a_N=0\mbox{  and  }b_N=P^{-1}\left(1-\frac{1}{N}\right)\ .
\ee
(c) Weibull
\be
a_N=x^\star\mbox{  and  }b_N=x^\star-P^{-1}\left(1-\frac{1}{N}\right)\ ,
\ee
where $x^\star$ as before is the upper endpoint of the support of $p(x)$.
\item Given a certain $p(x)$, it is possible to predict to which ``domain of attraction" (Gumbel, Fr\'echet or Weibull) its maximum belongs. Compute the following limit
\be
\lim_{\epsilon\to 0}\frac{P^{-1}(1-\epsilon)-P^{-1}(1-2\epsilon)}{P^{-1}(1-2\epsilon)-P^{-1}(1-4\epsilon)}=2^c\ .
\ee
If $c=0,>0,<0$, the domain of attraction is Gumbel, Fr\'echet or Weibull respectively.
\item The constants $\{a_N,b_N\}$ are not unique. If $\{a_N,b_N\}$ are suitable centering and scaling constants for a given $p(x)$, so are the constants $\{a^\prime_N,b^\prime_N\}$ provided the following limits hold
\begin{align}
\lim_{N\to\infty}\frac{b^\prime_N}{b_N}  &=1\label{cond1}\\
\lim_{N\to\infty}\frac{a_N-a^\prime_N}{b_N} &=0\label{cond2}\ .
\end{align}

\end{enumerate}
I also recommend the following references \cite{giorgy1,giorgy2} for an approach to EVS based on renormalization ideas and PDEs. 

I will mainly focus on the Gumbel universality class in the following.
In the next section, the exponential pdf will be used as a warm-up exercise to illustrate these basic notions, as well as the LDT treatment that was promised in the introduction.

\section{Warm-up: exponential pdf}
Consider for simplicity the case $p(x)=\mu\exp(-\mu x)$ for $x\geq 0$. 

\subsection{Limiting distribution}

From the general formalism
\be
Q_N(x)=\left[\mu\int_0^x \e^{-\mu y}\dd y\right]^N=\left[1-\e^{-\mu x}\right]^N=\e^{N\ln[1-\e^{-\mu x}]}\ .
\ee
As $x\to\infty$, expanding the logarithm one gets
\be
Q_N(x)\approx \e^{-N \e^{-\mu x}}=\e^{-\e^{-(\mu x-\ln N)}}\equiv F_{\mathrm{I}}(z)\ ,
\ee
if $z=\mu x-\ln N$. Here, $F_{\mathrm{I}}(z)$ is the Gumbel cdf. This implies that $a_N=\ln N/\mu$ and $b_N=1/\mu$. 

Of course, one could have derived them recalling \eqref{ANgumbel}. The cdf for the exponential pdf is
\be
P(x)=\mu\int_0^x\exp(-\mu y)\dd y=1-\e^{-\mu x}\ ,
\ee
hence $P^{-1}(x)=-(1/\mu)\ln (1-x)$ for $0\leq x<1$. Therefore
\be
a_N=P^{-1}\left(1-\frac{1}{N}\right)=\frac{\ln N}{\mu}\mbox{  and  }b_N=P^{-1}\left(1-\frac{1}{N\mathrm{e}}\right)-a_N=\frac{1}{\mu}\ ,
\ee
as expected. In the next section, I exploit a rare luxury offered by the exponential pdf: the distribution of the maximum can be computed also at \emph{finite} $N$. This offers the opportunity to understand at a somewhat deeper level the meaning of the centering and scaling constants $a_N$ and $b_N$.

\subsection{Finite $N$ and meaning of $a_N$ and $b_N$}

Let us compute the average and variance of $X_{\mathrm{max}}$ for \emph{finite} $N$. By definition

\begin{equation}
\mu_1(N) =\langle X_{\mathrm{max}}\rangle=\int_0^\infty \dd x\ x \frac{\dd}{\dd x}Q_N(x)=\frac{\gamma+\psi^{(0)}(N+1)}{\mu}\ ,
\end{equation}
where $\gamma=0.577216...$ is the Euler-Mascheroni constant, and $\psi^{(n)}(x)$ is the Polygamma function $(\psi^{(n)}(x)=\dd^n \psi(x)/\dd x^n$, where $\psi(x)=\Gamma^\prime(x)/\Gamma(x)$ is the logarithmic derivative of the Gamma function).

Taking the limit $N\to\infty$, we find

\begin{equation}
\mu_1(N) \sim \frac{\ln N}{\mu}+\frac{\gamma}{\mu}+\frac{1}{2\mu N}-\frac{1}{12\mu N^2}+\ldots,\qquad N\to\infty\ .\label{nu1}
\end{equation}

Now, one has $Q_N(a_N+b_N z)\to F_{\mathrm{I}}(z)$ for $N\to\infty$, where $F_{\mathrm{I}}(z)$ is the Gumbel cdf. The Gumbel pdf is
\be
f_{\mathrm{I}}(z)=\frac{\dd}{\dd z}F_{\mathrm{I}}(z)=\e^{-z-\e^{-z}}\ ,
\ee
whose average and variance can be computed as follows
\begin{align}
\int_{-\infty}^\infty\dd z\ z\ \e^{-z-\e^{-z}} &=\gamma\ ,\\
\sigma^2_G=\int_{-\infty}^\infty\dd z\ z^2\ \e^{-z-\e^{-z}} -\gamma^2 &=\frac{\pi^2}{6}\ .
\end{align}
The centering and scaling parameters $a_N$ and $b_N$ were computed in the last section as $a_N=\ln N/\mu$ and $b_N=1/\mu$.

Comparing with \eqref{nu1}, the leading term of the $N\to\infty$ expansion of the first moment turns out to be precisely equal to $a_N$, the centering parameter! This means that $a_N$ governs the average location of the maximum for large $N$. Moreover, the following holds
\be
\lim_{N\to\infty} \frac{\mu_1(N)-a_N}{b_N}=\lim_{N\to\infty} \frac{\mu_1(N)-\ln N/\mu}{1/\mu}=\gamma\ ,\label{explimitaN}
\ee
i.e. the average of the Gumbel pdf! Therefore the parameter $b_N$ ensures that the average location of the maximum in the large $N$ limit is adjusted to the (nonzero!) average of the limiting pdf (Gumbel).

Let us now compute the second moment and the variance of $X_{\mathrm{max}}$. One gets analogously:
\begin{equation}
\mu_2(N) =\langle X_{\mathrm{max}}^2\rangle=\int_0^\infty \dd x\ x^2 \frac{\dd}{\dd x}Q_N(x)=\frac{6
   H_N^2-6 \psi
   ^{(1)}(N+1)+\pi
   ^2}{6 \mu ^2}\ ,
\end{equation}
where $H_N=\sum_{k=1}^N 1/k$ is the $N$th harmonic number. Computing now the variance

\begin{equation}
\mathrm{Var}_{X_{\mathrm{max}}}(N)=\mu_2(N) -(\mu_1(N) )^2=\frac{\pi ^2-6\ \psi
   ^{(1)}(N+1)}{6 \mu ^2}\ ,
\end{equation}
which is exact for all $N$. Expanding for $N\to\infty$, we see that the variance saturates at a finite value, namely

\begin{equation}
\mathrm{Var}_{X_{\mathrm{max}}}(N)\sim \frac{\pi^2}{6}\frac{1}{\mu^2}-\frac{1}{\mu^2 N}+\ldots,\qquad N\to\infty\ .
\end{equation}
Interestingly, the saturating value $ \frac{\pi^2}{6}\frac{1}{\mu^2}$ has a quite natural interpretation as the product of i) the variance of the Gumbel pdf $\sigma^2_G=\pi^2/6$ and ii) the square of $b_N=1/\mu$ (the scaling parameter of the Extreme Value distribution). In summary

\begin{equation}
\lim_{N\to\infty} \frac{\mathrm{Var}_{X_{\mathrm{max}}}(N)}{b_N^2}=\sigma^2_G\ ,\label{explimitbN}
\end{equation}
implying that $b_N$ serves also the purpose of ``shrinking" the width of the pdf of $X_{\mathrm{max}}$ as much as needed to squeeze it under the envelope of the limiting (Gumbel) pdf for $N\to\infty$.

The properties \eqref{explimitaN} and \eqref{explimitbN} can be more compactly expressed using the notation $X_{\mathrm{max}}\sim a_N+b_N\chi$, with $\chi$ a Gumbel-distributed random variable. Standard properties (linearity and homogeneity) of cumulants then imply
$\langle X_{\mathrm{max}}\rangle\sim a_N+b_N\langle \chi\rangle$ (namely Eq. \eqref{explimitaN}) and $\mathrm{Var}_{X_{\mathrm{max}}}(N)=b_N^2 \mathrm{Var}(\chi)$ (namely Eq. \eqref{explimitbN}).

\subsection{Large deviations}

I set $\mu=1$ for simplicity in the following. So far I have considered the standard textbook treatment of EVS for exponential variates, which can be summarized in the statement
\be
 \mathrm{Prob}[X_{\mathrm{max}}>\ln N +z] \sim 1-\e^{-{\e^{-z}}}\ ,\qquad N\to\infty\ ,\label{statementlimiting}
\ee
with $z\sim\mathcal{O}(1)$ for large $N$. In the last subsection, $a_N=\ln N$ was shown to be the average location of the maximum in the large $N$-limit. Therefore, the ``classical" statement \eqref{statementlimiting} concerns \emph{typical} $\mathcal{O}(1)$ fluctuations around the average value in the large $N$ limit.

It is then natural to ask instead the following \emph{different} question: what is the probability that the maximum is ``much larger" than expected, meaning that it deviates from $\ln N$ (to the right) by an amount \emph{proportional} to $\ln N$?

In formulae,
\be
\mathrm{Prob}[X_{\mathrm{max}}>(\ln N)\xi]=?\qquad \xi\sim\mathcal{O}(1)\ .\label{question}
\ee
Note that this is evidently a rare event! Its probability must decay quite fast as $N$ increases. Still, it may not be completely clear at this stage whether the answer to the question in \eqref{question} is somehow implicitly contained already in the limiting statement \eqref{statementlimiting}. I will show later that this is \emph{not} the case: the large deviation results \emph{cannot} be in general deduced as a corollary of the limiting distribution \emph{alone}, which holds on a much narrower scale (for small fluctuations around the average). 

Computing \eqref{question} is rather straightforward
\begin{align}
\nonumber\mathrm{Prob}[X_{\mathrm{max}}>(\ln N)\xi] &=1-\mathrm{Prob}[X_{\mathrm{max}}\leq(\ln N)\xi]=1-Q_N((\ln N)\xi)\\
&=1-\left[1-\e^{-(\ln N)\xi}\right]^N\ .
\end{align}
The claim (immediate to verify using $(1-N^{-\xi})^N\sim 1-N^{1-\xi}$ for large $N$) is therefore
\be
\lim_{N\to\infty}\frac{-\ln \mathrm{Prob}[X_{\mathrm{max}}>(\ln N)\xi]}{\ln N}=
\begin{cases}
\xi-1 & \xi\geq 1\\
0 & \text{otherwise}
\end{cases}\ .\label{limitreally}
\ee
This simple result is expressed in a ``standard" large deviation form, albeit with a quite unusual speed $\ln N$, in contrast with the speeds $N$ and $N^2$ that are customary for i.i.d. sample means and random matrix observables \cite{majumdarreview}, respectively. The probability of a large fluctuation to the right of the expected value for the maximum, therefore, decays effectively as a power-law in $N$, 
$ \mathrm{Prob}[X_{\mathrm{max}}>(\ln N)\xi]\sim 1/N^{\xi-1}$,
with exponent given by the \emph{right rate function} $\psi_r(\xi)=\xi-1$. 

It is useful to summarize the two (small and large) deviation results presented so far
\begin{align}
 \mathrm{Prob}[X_{\mathrm{max}}>\ln N +z] &\sim 1-\e^{-{\e^{-z}}}\ ,\label{limit1}\\
\mathrm{Prob}[X_{\mathrm{max}}>(\ln N)\xi] &\approx \e^{-(\ln N)(\xi-1)}\ ,\label{limit3}
\end{align}
with $z$ and $\xi$ of $\mathcal{O}(1)$ for large $N$. 

I will show now that an interesting ``matching" occurs between the ``most unlikely" among typical fluctuations (probed by the limit $z\gg 1$ in \eqref{limit1}) and the ``most likely" among atypical fluctuations (probed by the limit $\xi\simeq 1$ in \eqref{limit3}).

For $z\gg 1$ one has
\be
 \mathrm{Prob}[X_{\mathrm{max}}>\ln N +z] \sim \e^{-z}\ .\label{limit2}
\ee
Setting now $\ln N+z\simeq (\ln N)\xi$, one obtains that in the matching regime $\xi\simeq 1+z/\ln N$, and substituting in \eqref{limit3}
\be
\e^{-(\ln N)(\xi-1)}\Big|_{\xi\simeq 1+z/\ln N}\simeq \e^{-z}\ ,
\ee
as in \eqref{limit2}. Hence, the large deviation \eqref{limit3} when approaching $\ln N$ ($\simeq 1$) from the right on a scale of $\mathcal{O}(1/\ln N)$ smoothly matches the far-right tail 
of the typical (limiting) distribution. I offer here two remarks, though:
\begin{enumerate}
\item While this matching property is naturally expected to hold, and it does in a few other examples I know \cite{majumdarreview}, it does not seem to have the status of a necessary/sufficient condition, encoded in a \emph{theorem} (at least, not that I am aware of). This would be a very interesting research direction to pursue, though, much in the spirit of Bryc's regularity condition for retrieving the Central Limit Theorem from the rate function \cite{bryc}.
\item 
Assuming that this matching must hold necessarily, it would have occurred for \emph{any} rate function $\psi_r(\xi)$ behaving as $\xi-1$ for $\xi\to 1$: therefore the \emph{true} rate function $\psi_r(\xi)$ (among all the possibilities) \emph{cannot} be deduced by this matching (i.e. by the behavior of the limiting distribution for $z\gg 1$) \emph{alone}: one really has to compute the limit \eqref{limitreally} ``from scratch"! 
This will be all the more evident in the next case.
\end{enumerate}

\section{Gaussian pdf}

An even more interesting case is the Gaussian pdf $p(x)=\e^{-x^2/2}/\sqrt{2\pi}$. We present in the following subsection a thorough derivation of the centering and scaling constants $a_N$ and $b_N$ for this case, as there are several subtleties that are worth highlighting.

\subsection{Limiting distribution}

From the general formalism
\be
Q_N(x)=\left[1-\int_x^\infty\dd y \frac{\e^{-y^2/2}}{\sqrt{2\pi}}\right]^N=\left[\frac{1}{2}(1+\mathrm{erf}(x/\sqrt{2}))\right]^N\ ,\label{QNgaussian}
\ee
where the error function $\mathrm{erf}(z)=(2/\sqrt{\pi})\int_0^z\dd t\ \e^{-t^2}$.

It is convenient to use the integral form for $Q_N(x)$ to derive the centering and scaling constants $a_N$ and $b_N$.
\be
Q_N(x)=\left[1-\int_x^\infty\dd y \frac{\e^{-y^2/2}}{\sqrt{2\pi}}\right]^N\simeq \exp\left[-N\int_x^\infty\dd y \frac{\e^{-y^2/2}}{\sqrt{2\pi}}\right]\ ,\label{Qngaussianexpanded}
\ee
as for $x\to\infty$ the integral gives a small contribution, and we can use $(1-\epsilon)^N\simeq \e^{-N\epsilon}$.

For $x\to\infty$, the behavior of the integral $I(x)=\int_x^\infty\dd y \e^{-y^2/2}$ can be estimated as follows. Make a change of variables $y=x\tau$, yielding
\be
I(x)=x\int_1^\infty \dd\tau\ \e^{-x^2\tau^2/2}\ .
\ee
The integrand is a fast decreasing function of $\tau$, so for large $x$ the main contribution to the integral comes from the vicinity of the point $\tau=1$. Expanding the function $\tau^2/2$ in the exponent close to $\tau=1$ as $\tau^2/2=1/2+1\times (\tau-1)+\ldots$
\be
I(x)\sim x\e^{-x^2/2}\int_1^\infty \dd\tau\ \e^{-x^2(\tau-1)}=\frac{\e^{-x^2/2}}{x}\ ,\qquad\text{for }x\to +\infty\ .\label{asymIx}
\ee
Inserting it in \eqref{Qngaussianexpanded}
\be
Q_N(x)\simeq \exp\left[-\frac{N}{\sqrt{2\pi}}\frac{\e^{-x^2/2}}{x}\right]=\e^{-\e^{-\varphi_N(x)}}\ ,\label{Qconvergegumbel}
\ee
where
\be
\varphi_N(x)=-\ln N+x^2/2+\ln(x)+(1/2)\ln(2\pi)\ .\label{varphi}
\ee
This looks quite promising in terms of convergence to the expected Gumbel form. Note that it is \emph{not} legitimate to drop the term $1/x$ with respect to $\e^{-x^2/2}$ in \eqref{asymIx} (or, equivalently, to drop the $\ln(x)$ and the constant in \eqref{varphi}) as one would be naively tempted to do.

Setting now $x=a_N+b_N z$ in \eqref{varphi} and expanding we obtain
\be
\varphi_N(a_N+b_N z)=-\ln N+\frac{1}{2}a_N^2+\frac{1}{2}b_N^2 z^2+a_N b_N z+\frac{1}{2}\ln(2\pi)+\ln(a_N+b_N z) \ .\label{varphi2first}
\ee
Imposing that \eqref{varphi2first} should go as $\sim z$ for $N\to\infty$ (as dictated by the Gumbel form $\e^{-\e^{-z}}$) gives the constraint
\be
b_N = \frac{1}{a_N}\ .
\ee
Next, in order to neutralize the term $-\ln N$, I put forward the ansatz $a_N=\sqrt{2 \ln N + c_N}$, obtaining
\be
\varphi_N\left(\sqrt{2 \ln N+ c_N}+\frac{z}{\sqrt{2 \ln N+ c_N}}\right)\simeq z+\frac{c_N}{2}+\frac{1}{2}\ln(2\pi)+\ln\left(\sqrt{2 \ln N+ c_N}+\frac{z}{\sqrt{2 \ln N+ c_N}}\right) \ ,\label{varphi2}
\ee
where I neglected the term $z^2/(2 a_N^2)$ which vanishes for $z\sim\mathcal{O}(1)$ and $a_N$ going to infinity when $N\to\infty$.

Expanding the last logarithm, and neglecting the term $z/(2\ln N+c_N)$ I obtain 
\begin{align}
\nonumber\varphi_N\left(\sqrt{2 \ln N+ c_N}+\frac{z}{\sqrt{2 \ln N+ c_N}}\right) &\simeq z+\frac{c_N}{2}+\frac{1}{2}\ln(2\pi)+\frac{1}{2}\ln(2\ln N+c_N)\\
&\simeq z+\frac{c_N}{2}+\frac{1}{2}\ln(2\pi)+\frac{1}{2}\ln (2\ln N)+\frac{1}{2}\frac{c_N}{2\ln N} \ .\label{varphi2}
\end{align}
The constant $c_N$ can now be determined by the condition
\be
\frac{c_N}{2}+\frac{1}{2}\ln(2\pi)+\frac{1}{2}\ln (2\ln N)+\frac{1}{2}\frac{c_N}{2\ln N}=0\Rightarrow c_N=\frac{-2 \ln N \ln (4\pi\ln N)}{1+2\ln N}\simeq -\ln(4\pi\ln N)\ .
\ee
In summary, the two centering and scaling constants for the Gaussian pdf are
\be
\begin{cases}
a_N &=\sqrt{2\ln N-\ln(4\pi\ln N)}\\
b_N &=\frac{1}{\sqrt{2\ln N-\ln(4\pi\ln N)}}\ .
\end{cases}
\ee
One can use the conditions \eqref{cond1} and \eqref{cond2} to simplify these expressions. The claim is that one can equivalently use\footnote{One often finds misprints in the (few) published resources where such constants are spelt out somewhat explicitly.}
\be
\begin{cases}
a^\prime_N &=\sqrt{2\ln N}-\frac{\ln (4\pi \ln N)}{2\sqrt{2\ln N}}\label{aNprime}\\
b^\prime_N &=\frac{1}{\sqrt{2\ln N}}\ .
\end{cases}
\ee
Indeed, the limits
\begin{align}
\lim_{N\to\infty}\frac{b^\prime_N}{b_N} &=\lim_{N\to\infty}\sqrt{\frac{2\ln N-\ln(4\pi\ln N)}{2 \ln N}}=1\\
\lim_{N\to\infty}\frac{a_N-a^\prime_N}{b_N}&= \lim_{N\to\infty}\frac{\sqrt{2\ln N-\ln(4\pi\ln N)}-\sqrt{2\ln N}+\frac{\ln (4\pi \ln N)}{2\sqrt{2\ln N}}}{1/\sqrt{2\ln N-\ln(4\pi\ln N)}}=0\ ,
\end{align}
as dictated by \eqref{cond1} and \eqref{cond2}.

I offer some remarks here.
\begin{enumerate}
\item It is \emph{not} legitimate to drop the second term in $a_N^\prime$ \eqref{aNprime} with the argument that it vanishes while the first diverges as $N\to\infty$. This would be tantamount to claiming that $a^{\prime\prime}_N=\sqrt{2\ln N}$ is equally fit to stand as centering constant. But the limit
\be
\lim_{N\to\infty}\frac{a_N-a^{\prime\prime}_N}{b_N}=\lim_{N\to\infty}\frac{\sqrt{2\ln N-\ln(4\pi\ln N)}-\sqrt{2\ln N}}{1/\sqrt{2\ln N-\ln(4\pi\ln N)}}=-\infty\ ,
\ee
in violation of the requirement \eqref{cond2}.

\item The correctness of the constants \eqref{aNprime} can also be ascertained numerically. \textsf{Mathematica} is able to compute the limit $\lim_{N\to\infty}Q_N(a^\prime_N+b^\prime_N z)$ for a specific value of $z$. Starting from \eqref{QNgaussian}, I write the following two lines of code
\begin{verbatim}
Q[NN_, z_] := ((1/2) (1 + Erf[(Sqrt[2 Log[NN]] 
- Log[4 Pi Log[NN]]/(2 Sqrt[2 Log[NN]]) + z/Sqrt[2 Log[NN]])/Sqrt[2]]))^NN;
Limit[Q[NN, 0.001], NN -> Infinity]           
>> 0.368247
Exp[-Exp[-0.001]]
>> 0.368247
\end{verbatim}
Similarly, one can further disprove the naive use of $a^{\prime\prime}_N=\sqrt{2\ln N}$ as a centering constant with the code
\begin{verbatim}
Qfalse[NN_, z_] := ((1/2) (1 + Erf[(Sqrt[2 Log[NN]] 
+ z/Sqrt[2 Log[NN]])/Sqrt[2]]))^NN;
Limit[Qfalse[NN, 0.001], NN -> Infinity]           
>> 1.
Exp[-Exp[-0.001]]
>> 0.368257
\end{verbatim}

\end{enumerate}

\subsection{Large deviations}
As for the exponential pdf, I now wish to address the probability of anomalously large fluctuations of the maximum to the right of the expected value (the centering constant $a_N$ or $a_N^\prime$). More precisely, I wish to compute
\be
\mathrm{Prob}[X_{\mathrm{max}}>a^\prime_N \xi]=?
\ee 
and how this probability decays for large $N$.

The calculation can be performed easily
\be
\mathrm{Prob}[X_{\mathrm{max}}>a^\prime_N \xi]=1-Q_N(a^\prime_N\xi)=1-\left[1-\int_{a^\prime_N\xi}^\infty\frac{\e^{-y^2/2}}{\sqrt{2\pi}}\dd y\right]^N\simeq N\int_{a^\prime_N\xi}^\infty\frac{\e^{-y^2/2}}{\sqrt{2\pi}}\dd y\ .
\ee
The integral can be estimated in full analogy with $I(x)$ in \eqref{asymIx}, yielding
\be
\mathrm{Prob}[X_{\mathrm{max}}>a^\prime_N \xi]\simeq \frac{N\e^{-a^{\prime 2}_N \xi^2/2}}{\sqrt{2\pi}\ a^\prime_N \xi}\ .
\ee
Taking the logarithm on both sides, dividing by $\ln N$ (the \emph{speed}) and replacing the definition of $a^\prime_N$ from \eqref{aNprime}, one obtains the formidable limit
\begin{align}
\nonumber &\lim_{N\to\infty}\frac{-\ln \mathrm{Prob}\left[X_{\mathrm{max}}>\left(\sqrt{2\ln N}-\frac{\ln (4\pi \ln N)}{2\sqrt{2\ln N}}\right) \xi\right]}{\ln N}=\\
&\lim_{N\to\infty}-\frac{1}{\ln N}\ln \left(\frac{N \exp \left(-\frac{1}{2} \xi ^2 \left(\sqrt{2\ln N}-\frac{\ln (4 \pi  \ln N)}{2 \sqrt{2\ln N}}\right)^2\right)}{\sqrt{2 \pi } \xi  \left(\sqrt{2\ln N}-\frac{\ln (4 \pi  \ln N)}{2 \sqrt{2\ln N}}\right)}\right)=\xi^2-1\equiv \psi_r(\xi),\qquad \xi\geq 1\ .
\end{align}

Again, the two (small and large) deviation results can be summarized as
\begin{align}
 \mathrm{Prob}\left[X_{\mathrm{max}}>a^\prime_N +b^\prime_N z\right] &\sim 1-\e^{-{\e^{-z}}}\ ,\label{limit1gauss}\\
\mathrm{Prob}\left[X_{\mathrm{max}}>a^\prime_N \xi\right] &\approx \e^{-(\ln N)(\xi^2-1)}\ ,\label{limit3gauss}
\end{align}
with $z$ and $\xi$ of $\mathcal{O}(1)$ for large $N$. The constants $a^\prime_N$ and $b^\prime_N$ are given in \eqref{aNprime}.

Expanding the rate function $\psi_r(\xi)=\xi^2-1$ around $\xi=1$, one obtains
\be
\psi_r(\xi)\sim 2(\xi-1)\ .\label{psitaylor}
\ee

For $z\gg 1$ one has
\be
 \mathrm{Prob}\left[X_{\mathrm{max}}>a^\prime_N +b^\prime_N z\right] \sim \e^{-z}\ .\label{limit2gauss}
\ee
Setting now $a^\prime_N+b^\prime_N z\simeq a^\prime_N\xi$, one obtains that in the matching regime $\xi\simeq 1+b^\prime_N z/a^\prime_N\simeq 1+z/(2\ln N)$, and substituting in \eqref{limit3gauss} with the expanded rate function \eqref{psitaylor}
\be
\e^{-2(\ln N)(\xi-1)}\Big|_{\xi\simeq 1+z/(2\ln N)}\simeq \e^{-z}\ ,
\ee
as in \eqref{limit2gauss}. Hence, once again the large deviation \eqref{limit3gauss} when approaching $\ln N$ ($\simeq 1$) from the right on a scale of $\mathcal{O}(1/\ln N)$ smoothly matches the far-right tail 
of the typical (limiting) distribution.

This example further confirms that in any case the full rate function $\psi_r(\xi)=\xi^2-1$ could not have been predicted appealing to the matching property alone, which only requires that the expansion around $\xi=1$ is $\approx 2(\xi-1)$. Therefore the large deviations results \eqref{limit3} and \eqref{limit3gauss} genuinely provide extra information, which is not carried by the limiting distribution (Gumbel) alone.

It is also easy to deduce along the same lines that the rate function $\psi_r(\xi)$ for the maximum of i.i.d. variables whose common pdf decays at infinity as $p(x)\sim \e^{-x^\delta}$ is given by $\psi_r(\xi)=\xi^\delta-1$, in agreement with \cite{giuliano}. I am not aware of a similar LDT treatment for EVS of densities in the Fr\'echet basin of attraction, while for the Weibull class this is also possible (with speed $N$), but somewhat much less interesting \cite{giuliano}.

\section{Conclusions}
In summary, having in mind an audience of theoretical and statistical physicists, I have presented some aspects of Extreme Value Statistics (restricted to the Gumbel basin of attraction) from the somewhat non-standard viewpoint of Large Deviation Theory. First, an introduction to the universality classes for the EVS statistics was given, and then the exponential and Gaussian parent pdf were worked out in detail. I pointed out some subtleties connected with the centering and scaling constants $a_N$ and $b_N$ for the Gaussian case, which are difficult to find discussed in the literature, and eventually the right rate function $\psi_r(\xi)$ was derived for large deviations of the maximum to the right of its expected value in both cases. Demonstrating a smooth matching between the far tail of the limiting distribution (small deviation) and the large deviation result, I stressed that the rate function cannot be deduced from the knowledge of the limiting distribution (Gumbel) alone, thus implying that it carries additional information. This will not come as a surprise for the reader familiar with the LDT for the \emph{maximum} eigenvalue of $N\times N$ Gaussian and Wishart random matrices \cite{majumdarreview}: the role of Gumbel is taken by the Tracy-Widom distribution there \cite{tracy1,tracy2}, while right and left rate functions (corresponding to speeds $N$ and $N^2$, respectively) were independently derived using different strategies \cite{dean,dean2,vergassola}. The corresponding (much simpler) result for i.i.d. variables, which is at the same time cute and instructive, seems to deserve a better fate than the oblivion it has fallen into.

\vspace{10pt}
{\bf Acknowledgments:} I am indebted with Fabio Caccioli, Fabio D Cunden, Giacomo Livan, Alberto Rosso, Pierfrancesco Urbani and Dario Villamaina for a careful reading of the manuscript and many useful suggestions, and to Satya N Majumdar for a clarifying correspondence. I am also genuinely grateful to two anonymous referees who offered valuable advice and prompted me to write a clearer paper. I acknowledge support from EPSRC Centre for Doctoral Training in Cross-Disciplinary Approaches to Non-Equilibrium Systems (CANES).

\end{document}